# Summarization and Matching of Density-Based Clusters in Streaming Environments[*]

Di Yang[†]
Oracle Corporation
1 Oracle Drive
Nashua, NH, USA
di.yang@oracle.com

Elke A. Rundensteiner
Worcester Polytechnic Institute
100 Institute Road
Worcester, MA, USA
rundenst@cs.wpi.edu

Matthew O. Ward
Worcester Polytechnic Institute
100 Institute Road
Worcester, MA, USA
matt@cs.wpi.edu

## ABSTRACT

Density-based cluster mining is known to serve a broad range of applications ranging from stock trade analysis to moving object monitoring. Although methods for efficient extraction of density-based clusters have been studied in the literature, the problem of summarizing and matching of such clusters with arbitrary shapes and complex cluster structures remains unsolved. Therefore, the goal of our work is to extend the state-of-art of density-based cluster mining in streams from cluster extraction only to now also support analysis and management of the extracted clusters. Our work solves three major technical challenges. First, we propose a novel multi-resolution cluster summarization method, called Skeletal Grid Summarization (SGS), which captures the key features of density-based clusters, covering both their external shape and internal cluster structures. Second, in order to summarize the extracted clusters in real-time, we present an integrated computation strategy C-SGS, which piggybacks the generation of cluster summarizations within the online clustering process. Lastly, we design a mechanism to efficiently execute cluster matching queries, which identify similar clusters for given cluster of analyst's interest from clusters extracted earlier in the stream history. Our experimental study using real streaming data shows the clear superiority of our proposed methods in both efficiency and effectiveness for cluster summarization and cluster matching queries to other potential alternatives.

## 1. INTRODUCTION

**Motivation.** Mining complex patterns such as clusters and graphs from huge volumes of streaming data has been recognized as critical for numerous application domains. To facilitate such complex pattern mining process, a streaming pattern mining system does not only need to be equipped with highly efficient pattern extraction algorithms, but more importantly, it must also provide effective pattern analysis support, as motivated below:

1) **Pattern feature abstraction.** The key features of detected patterns may be complex and thus may not be easily comprehensible for human analysts without analytical assistance. For example, in real-time traffic monitoring, a cluster representing a congestion area in the traffic of Beijing may be composed of 10K or even more vehicles and may spread to over $10km^2$. By simply looking at the information about individual cluster members (vehicles), such as their positions and moving speed, an analyst may not be able to identify the key features of this cluster in real time, such as where is the key bottleneck causing the congestion.

2) **Pattern compression.** Some patterns need to be kept for long-term analysis, yet keeping the full representation of the complex patterns tends to be impractical in streaming environments. In the previous example, storing the full representation of the detected traffic congestion patterns (arbitrarily shaped clusters), namely the individual cluster member tuples (tens of thousands tuples for each cluster) would cause not only a huge burden on the storage space but also low efficiency for pattern transmission.

3) **Pattern retrieval (matching).** For stream analysis, the archived patterns may need to be retrieved based on their features. Using the above example, when a new traffic congestion arises, the analysts may ask whether similar congestion patterns have been detected before. If yes, rather than figuring out a new congestion-relief plan from scratch, the previous proven-to-work solution for such congestion patterns could be directly applied.

In short, an effective pattern summarization method is the key for complex pattern analysis and management. It is needed for many different aspects of pattern analysis, including feature abstraction, compression and pattern retrieval (as mentioned above). Also, the pattern summarizations can also be used for approximated pattern representation. For example, one can design pattern visualization or full representation re-generation techniques based on pattern summarizations. In this work, our goal is to design effective **summarization** and **matching** techniques for density-based clusters in streaming environments, which remain open problems for database community.

**Sliding Window Semantics.** In this work, we focus on density-based cluster mining in *sliding stream windows* [7, 8, 16, 17]. In this query semantics, arbitrarily shaped clusters are continuously detected within the most recent portion of the stream. The traffic congestion monitoring task discussed above is an example that requires such query semantics. Other applications that require such query semantics include detecting intensive-transaction areas (clusters) in most recent stock trades, and identifying malicious attacks (clusters) in current network traffic.

---

[*]This work is supported by the NSF, under grants CCF-0811510, IIS 0812027 and IIS 1018443

[†]This work is done when the author is working at WPI.

**Challenges.** Summarization and matching of density-based clusters is not only an unsolved but also a challenging problem. To serve real-time streaming applications, the proposed techniques must address the following challenges: 1) Cluster summarization must be sufficiently **descriptive** yet highly **compact**. The cluster structure of a density-based cluster is defined by a series of densely populated sub-regions and as well as the connections among them (See Figure 1). Clearly, simple statistical aggregations, such as the centroid or minimum bounding rectangle of a cluster, are insufficient for describing such complex pattern structure. 2) The cluster summarization process has to be highly **efficient**. A system conducting expensive online clustering can hardly afford additional system resources for summarizing clusters in real-time. 3) The summarized cluster representation needs to be effectively **retrievable** ("matchable"). The matching process between cluster summarizations ought to loyally reflect the similarity between the original clusters, yet be computationally efficient.

**Proposed Solution.** To address the above challenges, we first analyze density-based cluster structures and identify their key characteristics, namely *position*, *shape*, *connectivity* and *density distribution*. To capture these features, we investigate two commonly-used summarization principles, namely the graph-based and the grid-based strategies, We discover that neither of them alone is capable to provide an effective summarization for density-based clusters. Therefore, we propose a hybrid solution, called Skeletal Grid Summarization (SGS). For descriptive power, SGS is shown to guarantee its fidelity to the original clusters on all key features. For compactness, our experimental study in Section 8 confirms that even the SGS of the highest resolution achieves on average a 98% compression rate of the full representation of the clusters.

Empowered by the proposed SGS summarization, we design a framework to support both continuous cluster extraction and cluster matching queries. A continuous cluster extraction query in our system does not only extract clusters in their full representation (all cluster member objects) for online monitoring purposes like the other state-of-the-art techniques [3, 16], but it also concurrently compacts them into the SGS summarization. The full and the summarized (SGS) representation formats are complementary to each other, providing a description of the clusters at the individual tuple and cluster feature level respectively. To extract these two representation formats simultaneously and in a highly efficient manner, we propose an integrated cluster extraction + summarization algorithm, C-SGS. C-SGS incrementally maintains both the full representation and the corresponding SGS of the extracted clusters in an integrated manner. This results in an almost "free" cluster summarization generation by piggy-packing the summarization process into the cluster extraction process itself. Our experimental study in Section 8 shows that C-SGS, which returns clusters in both full and summarized representation (SGS), has a neglectable overhead, compared with state-of-the-art algorithm Extra-N [16] computing the full representation of clusters only. In all our test cases, the extra response time of C-SGS compared with Extra-N is consistently less than 6% (Section 8.1).

For any "to-be-matched" cluster specified by the analyst, a cluster matching query identifies similar clusters extracted earlier in the same stream from a pattern archive. To support such queries, our framework first archives the SGS of the extracted clusters into a pattern archive. When executing a cluster matching query, our system deploys a filter-and-refine strategy. First, the filter-phase exploits a feature index to locate the potential matching candidates from the pattern store. Then, the refine-phase conducts a more detailed cluster match against these promising candidates and returns those with similarity above a given threshold. Our experimental study shows that, efficiency-wise, our system takes only 3 seconds on average to answer a cluster matching query against 10K archived clusters (Section 8.2). Quality-wise, our user study, which invites human analysts to visually compare the similarity between matched clusters, shows that human analysts agree with a significant larger percentage of the matched clusters found using our proposed matching mechanism compared to those found by alternatives (Section 8.3).

**Contributions.** The main contributions of this work include: 1) We propose the first summarization method specifically designed for density-based clusters, namely the Skeletal Grid Summarization (SGS), 2) We present an integrated cluster mining and summarization algorithm, C-SGS, which efficiently computes the full representation and the SGS of the extracted clusters in one shot. 3) We develop a cluster matching mechanism based on SGS to efficiently processing cluster matching queries in real-time. 4) Our performance evaluation and user study using real streaming data confirm that our proposed techniques are clearly superior to other alternatives in all aspects, including summarization efficiency, cluster matching efficiency and matching quality.

## 2. RELATED WORK

The concept of density-based clustering was first proposed in [8]. It has drawn significant research attention [7, 16, 17, 12, 3, 4], because of its capability of identifying clusters with arbitrary shapes and specified density. Previous work mainly studied how to efficiently extract such clusters in static [8, 7, 12] or streaming environments [16, 17, 3, 4]. Also, given the prevalence of real-time monitoring tasks in stream applications, researchers have started to design visual platforms allowing human analysts to interactively explore such patterns in streams [14].

However, the fundamental problem of summarizing this important pattern type has not been studied in the literature yet. Without an effective yet compact summarization method, each density-based cluster has to be expressed by its full representation, namely its cluster member objects. Obviously, such full representation is neither succinct nor does it explicitly reflect the features of each cluster. This causes serious inconvenience for both storage and analysis of density-based clusters.

Traditional clustering methods [10, 19], such as k-mean style clustering, treat clusters as statistical phenomena. Therefore, many key features of the clusters, such as their shapes and densities, are summarized using a rather simplistic description. In particular, first, these works assume clus-

122

ters are spherically shaped. Therefore, the shape of a cluster is usually described using a simple "$centroid + radius$" formula. Second, the previous work do not capture the internal features of the clusters, such as how its density is distributed. For example, the density of a cluster is either treated as uniform or varying along the radius only. Obviously, such simple formula cannot well describe the complex cluster structure of density-based clusters. This is because both the shapes and density distributions of density-based clusters can be arbitrary, not to mention the complex sub-region connectivities in each cluster. To the best of our knowledge, no summarization method has been specifically designed for density-based clusters.

For computing cluster summarizations in streaming environments, if the clusters are treated as statistical phenomena, they are considered to be "aggregatable" over time [1, 5]. For example, [1] used one *Cluster Feature Vector* (CFV) to represent each micro-cluster detected in the stream. They rely on the *additivity property* of the CFV to aggregate the cluster features over time and compare the features of a same cluster at different time points by subtracting its CFVs on the corresponding time points.

However, the complex cluster structure of density-based clusters is not simply aggregatable over the sliding windows. The continuous expiration of old objects and arrival of new objects at each window may cause complex cluster structural changes, such as merge and split and connectivity changes within the clusters. Clearly, these changes cannot be simply captured by aggregation results. Thus, these techniques cannot effectively capture the features of density-based clusters within sliding windows.

## 3. PRELIMINARIES

### 3.1 Density-Based Clustering in Windows

Density-based cluster detection [8, 7] uses a range threshold $\theta^r \geq 0$ to define the neighbor relationship between objects. For two objects $p_i$ and $p_j$, if the distance between them is no larger than $\theta^r$, $p_i$ and $p_j$ are said to be neighbors. We use the function $NumNeigh(p_i, \theta^r)$ to denote the number of neighbors a object $p_i$ has, given the $\theta^r$ threshold.

**Definition** 3.1. *Density-Based Cluster: Given $\theta^r$ and a count threshold $\theta^c$, an object $p_i$ with $NumNeigh(p_i, \theta^r) \geq \theta^c$ is defined as a core point. Otherwise, if $p_i$ is a neighbor of any core object, $p_i$ is an edge point. $p_i$ is a noise point if it is neither a core object nor an edge object. Two core objects $p_0$ and $p_n$ are connected, if they are neighbors of each other, or there exists a sequence of core points $p_0, p_1, ... p_{n-1}, p_n$, where for any i with $0 \leq i \leq n-1$, each pair of core points $p_i$ and $p_{i+1}$ are neighbors of each other. Finally, a density-based cluster is defined as a maximum group of "connected core objects" and the edge objects attached to them. Any pair of core objects within a cluster are "connected".*

Figure 1 shows an example of a density-based cluster composed of 11 *core objects* (black) and 24 *edge points* (grey).

We focus on periodic sliding window semantics as proposed in CQL [2] and widely used in the literature [16, 17]. These proposed semantics can be either time- or count-based. Each query has a window with a fixed window size *win* and a fixed slide size *slide* (either a time interval or a tuple count). Clusters are generated for each window $W_i$ only based on those data points that fall into the same window $W_i$. Each cluster is returned as all its cluster member objects associated with the same cluster identification. We call this typical output format the **full representation** of each cluster.

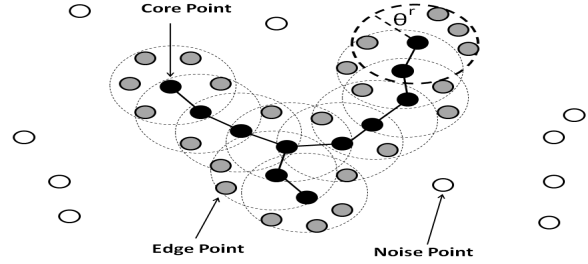

**Figure 1: Definition of Density-Based Clusters**

### 3.2 Supported Queries and System Overview

Our system support two types of analytical queries:

**Continuous Clustering Queries.** A *Continuous Custering Query* returns both *full* (Section 3.1) and *summarized representation* of the extracted clusters (Figure 2). The design of our proposed cluster summarization format will be introduced in Section 4.

---
**DETECT** $DensityBasedClusters^{f+s}$ **FROM** $stream$
**USING** $\theta^{range} = r$ **and** $\theta^{cnt} = c$
**IN Windows WITH** $win = w$ **and** $slide = s$

---

**Figure 2: Continuous Cluster Extraction Query Returning full *(f)* and summarized *(s)* representations of clusters**

**Cluster Matching Queries.** Given a user specified to-be-matched cluster $C_i$, a cluster matching query finds clusters similar to $C_i$ that reside in the historical pattern archive. We show a template of such a query in Figure 3.

---
**GIVEN** $DensityBasedCluster^s$ $C_i$
**SELECT** $DensityBasedCluster^s$ $C_j$ **FROM** $History$
**WHERE** $Distance(C_i, C_j) \leq sim\_threshold$

---

**Figure 3: Cluster Matching Query finding Clusters Similar to To-Be-Matched Cluster Based on Cluster Summarization**

The to-be-matched cluster can be any cluster specified by an analyst. Typically, it may be a cluster detected in the most recent portion of the stream that represent the newest characteristics of the stream. The matched clusters, if any, will be found in the historical pattern store, which archives the clusters extracted by *Continuous Clustering Query* earlier in the stream.

### 3.3 System Overview

To support these two types of analytical queries, we design a framework composed of four major components (Figure 4). Here we give a brief overview of the functionalities of each

123

component, while in-depth technical details are discussed later in Sections 5 to 7.

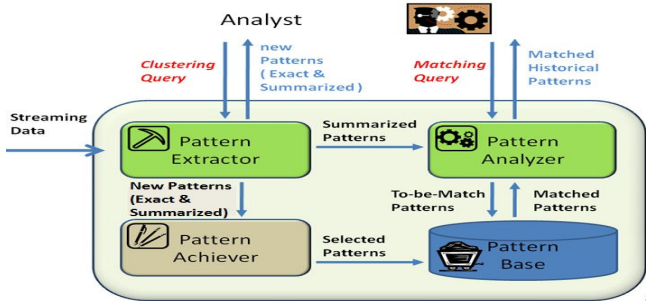

Figure 4: System Overview

The **Pattern Extractor** executes the *Continuous Cluster Extraction Query* (Figure 2) against the input stream. It outputs both *full* and *summarized representations* of the extracted clusters. Both representations are returned to the analyst for real-time monitoring. Meanwhile, the extracted clusters are also passed to the Pattern Archiver for storage, and to Pattern Analyzer for cluster matching.

The **Pattern Archiver** selectively archives the newly detected clusters into the Pattern Base. These archived clusters constitute the *Stream History* available for subsequent *Cluster Matching Queries* (Figure 3). The Pattern Archiver controls which extracted clusters should be kept in the Pattern Base and at which resolution they should be archived.

The **Pattern Base** organizes the archived clusters. To facilitate cluster matching against historical clusters, it employs multiple feature indices to organize the archived clusters. This helps the *Cluster Matching Queries* to quickly locate the potential matching candidates.

The **Pattern Analyzer** executes the *Cluster Matching Queries* (Figure 3). If an analyst is interested in any newly extracted cluster and would like to learn whether similar clusters had been detected before in the *Stream History*, she can submit her *Cluster Matching Query* to the Pattern Analyzer to search for matches against the Pattern Base.

## 4. CLUSTER SUMMARIZATION

### 4.1 Features of Density-Based Clusters

Based on our analysis, we identify four key features that define each density-based cluster, which can be divided into two categories, namely external and internal features.

**External Features:**

*Location:* The location of a cluster indicates its position in the data space. It provides basic information about each cluster, such as where a congestion area (a cluster) arises in the traffic, or in which price range an intensive-transaction area, a cluster based on price, volume and transaction time, is detected in the stock transaction stream.

*Shape:* Density-based clusters can have arbitrary shapes. The shape is a key feature, because a certain shape of the cluster may convey specific meaning for an application. For example, for the clusters representing intensive-transaction areas in stock transactions, a cluster having a long spread on transaction price but short range on transaction time conveys that a large number of transactions of a certain stock happened in a short time period while the price of it fluctuated dramatically within this time period.

**Internal Features:**

*Connectivity:* The connectivity of a density-based cluster describes how sub-regions within the cluster are connected. It is important for density-based clusters for both definition and application reasons. First, it defines internal structure of each cluster. The definition of the density-based cluster (see Section 3.1) relies on the connectivities among sub-regions to define a cluster. Second, the connectivies among sub-regions may be relevant to applications. For example, if two sub-regions within a single cluster representing a group of moving troops are not directly connected, then this may indicate the units in these two sub-regions cannot directly communicate with each other, because there are no connected "Head Nodes" (core objects) in these two sub-regions of their wireless network.

*Density Distribution:* Although the definition of density-based clusters imposes a minimal density requirement on objects in a cluster, the density of each cluster can be rather diverse across its sub-regions. The density distribution within each cluster may be of an analyst's interest in many applications. Using the earlier example, even in a single congestion area, the level of congestion (density of vehicles) may vary among sub-regions. Therefore, the density distribution in each sub-region may be the key for working out a congestion relief plan, as the super dense sub-regions may be the areas that cause the congestion.

### 4.2 Initial Effort: Graph-Based Summarization Method

Any effective *summarized representation* for density-based clusters has to capture the above four key features (Section 4.1). Given that density-based clusters may vary arbitrarily in shape, connectivity and also density distributions, using any aggregative method to represent these features will have rather poor descriptive power. Therefore, we propose to leverage an alternative strategy, namely the divide-and-conquer approach. We divide each cluster into sub-regions, and then we describe not only the features in each sub-region but also the interrelationships among the sub-regions.

Given this divide-and-conquer strategy, we first introduce a possible summarization method based on graph theory. This method uses one representative object to represent each sub-region. We call it "Skeletal Point Summation" (SkPS):

**Definition** 4.1. *For each cluster $C_i$, the SkPS summarization of $C_i$ is a graph $G(V, E)$ composed of a minimal set of connected core objects of $C_i$, called Skeletal Points as vertices $V$, whose neighborhoods together cover all the objects in this cluster, and connections among them as edges $E$.*

The graph composed of all *core objects* in Figure 1 is an example for SkPS. SkPS captures most of the cluster features and also has good compactness. However, it suffers from several serious shortcomings. First, SkPS has limited descriptive power for a cluster's density distribution. Second, such SkPS is not efficiently computable. For each cluster, identifying its SkPS is equal to the problem of identifying the **connected dominant set** in an undirected graph which has been proven to be NP-complete [9]. Third, SkPS is not a viable solution for matching, because a single cluster may have multiple SkPSs with rather different graph structures. Based on our analysis, these limitations suffered by SkPS are caused by its overlapping and non-deterministic sub-region division strategy. In conclusion, SkPS does not constitute an ideal summarization for density-based clusters. A more detailed discussion of SkPS method can be found in our technical report [18].

## 4.3 Proposed Solution: Skeletal Grid Summarization Method

**Basics of Grid-Based Summarization.** To solve the limitations suffered by SkPS, we propose to adapt SkPS by dividing each cluster into non-overlapping sub-regions. In particular, we divide the whole data space into uniformly sized grid cells. For each cluster, its sub-region division is now determined by the grid cells into which its members fall. Therefore, a cluster $C_i$ can be represented by all the grid cells containing at least one of $C_i's$ cluster member objects.

**Connectivity Preservation.** However, this simplistic grid-based summarization lacks one key capability of the SkPS solution, namely it does not capture the connectivity within clusters. In SkPS, both the inner and inter sub-region connectivity information of each cluster is well preserved. First, each sub-region in SkPS itself is "well connected", as all objects in a sub-region are neighbors of the same *skeletal point*. Second, the inter connections among different sub-regions are explicitly expressed by the "edges" in SkPS. While this simplistic grid-based summarization preserve neither of these two types of connectivity information.

**Connectivities In Grid Cells.** To solve this problem, we propose to integrate the concept of "connectivities" into the grid-based solution. As foundation, we first introduce the concept of *status* to a grid cell. We divide the grid cells in each cluster's summarization into two categories, namely "*core cells*" and "*edge cells*".

**Definition** 4.2. ***Core cells***: a core cell of a cluster $C_i$ contains at least one core object (See Def. 3.1) of $C_i$.

***Edge cells***: an edge cell of a cluster $C_i$ contains no core object, but at least one edge object (See Def. 3.1) of $C_i$.

***Noise cells***: a noise cell contains neither core nor edge objects of any cluster. [1]

For **inner-sub-region connections**, we follow the basic principle for the sub-region division strategy, which is to pursue homogeneity in each sub-region. In particular, we pick a fine grid size to guarantee that the objects that fall into the same grid cell are neighbors of each other. More precisely, the diagonal of each grid is set to be equal to the range threshold $\theta^r$ in the given clustering query (see Section 3.1). This grid cell size selection will be relaxed later in our discussion of the multi-resolution cluster summarization (Section 6). Under this fine grid size selection, the *core* and *edge cells* can be shown to have the following properties.

**Lemma** 4.1. *All objects in a core cell belong to the same cluster.*

*Proof:* Since each *core cell* contains at least one *core object* and all the objects in each *core cell* are now neighbors of each other, it implies that all objects in the same *core cell* are neighbors of at least one common *core object*. Based on the definition of density-based cluster (see Def. 3.1), the neighbors of a *core object* belong to the same cluster. ∎

---
[1] *noise grid* are are only used in cluster computation stage.

**Lemma** 4.2. *The number of objects in an edge cell must be less than the count threshold $\theta^c$ in the clustering query.*

*Proof:* We prove this lemma by contradiction. Given that all objects in a grid cell are neighbors of each other, if there are at least $\theta^c$ objects in an *edge cell*, those objects would be *core objects*, as they all have at least $\theta^c$ neighbors. This contradicts the definition of *edge grid* (Def. 4.2). ∎

Given these properties, each grid cell is "well-connected" and constitutes a basic unit for the inter-grid connection expression, as defined below.

For the **inter-sub-region connection**, we now define the "connections" between grid cells.

**Definition** 4.3. *Two core cells $ccl_1$ and $ccl_2$ are **directly connected**, if there exists at least one core object $p_i$ in $ccl_1$ and one core object $p_j$ in $ccl_2$ that are neighbors of each other. Two core cells $ccl_0$ and $ccl_n$ are **connected**, if they are directly connected to each other, or there exists a sequence of core cells $ccl_0, ccl_1, ...ccl_{n-1}, ccl_n$, where for any $i$ with $0 \leq i \leq n-1$, each pair of core cells $ccl_i$ and $ccl_{i+1}$ are directly connected with each other.*

*An edge cell $ecl_i$ is **attached** to a core grid $ccl_j$, if there exists at least one object $p_i$ in $ecl_i$ and one core object $p_j$ in $ccl_j$ that are neighbors of each other.*

*Two edge cells are neither connected nor attached.*

Given the connection definition for grid cells above, all *core cells* of a cluster $C_i$ are *connected* to each other, and all *edge cells* are *attached* to at least one *core cell* of $C_i$.

**Skeletal Grid Summarization.** Based on the status and connections of grid cells, we now give the definition of our proposed Skeletal Grid Summarization (SGS) method.

**Definition** 4.4. *A **Skeletal Grid Summarization** (SGS) of a density-based cluster $C_i$ is composed of all grid cells that contain at least one cluster member object of $C_i$. We call each grid cell in a SGS, a **Skeletal Grid Cell** (Sc) of $C_i$. $SGS = \{Sc_0, Sc_1, ...Sc_n\}$. Each $Sc_i$ has five attributes, namely $SG_i =$*
*(location[], sidelength, population, status, connection[]).*

*1) location vector: a sequence of values, each indicating the minimum value on one of the dimensions covered by $Sc_i$.*

*2) side length: the range of values on each dimension.*

*3) population: the number of objects contained by $Sc_i$*

*4) status: whether $Sc_i$ is a core or edge cell.*

*5) connection vector: a sequence of boolean connection indicators, each indicating $Sc_i$'s connection to one of its adjacent skeletal grid cells. For any edge or noise cell, all connection indicators are "false". For any core grid, a connection indicator is "true" if the corresponding adjacent skeletal grid cell $Sc_j$ is a core cell and $Sc_i$ and $SG_j$ are directly connected, or if $SG_j$ is an edge cell attached to $SG_i$.*

Figure 5 shows an example of our proposed Skeletal Grid Summarization (SGS) for a 2D cluster. SGS achieves our goal of preserving all four features, as shown below.

**Lemma** 4.3. ***Fidelity to Location and Shape:*** *The data space covered by $C_i.SGS$ is larger than that covered by the cluster member objects of $C_i$ by a bounded error. Namely, any point in the data space covered by $C_i.SGS$ is at most $\theta^r$ away from a cluster member object in $C_i$.*

*Proof:* The data space covered by $C_i.SGS$ is composed of the union of the space covered by all its *skeletal grid cells*.

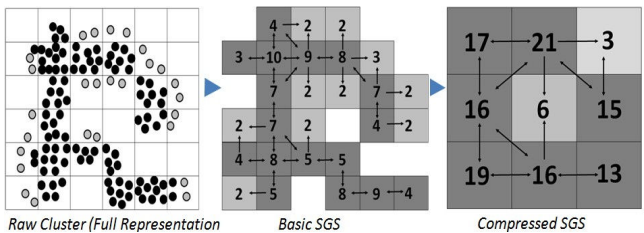

Figure 5: Example of full representation, basic SGS and compressed SGS of a 2D cluster

Since all member objects of $C_i$ fall into these grid cells, the data space covered by $C_i.SGS$ is larger than that covered by $C_i$'s member objects. Since each skeletal grid cell in $C_i.SGS$ contains at least one member of $C_i$, and the diagonal of each cell is $\theta^r$, any point in the data space covered by a skeletal grid cell is at most $\theta^r$ away from a member of $C_i$. ∎

**Lemma** 4.4. *Fidelity to Density Distribution:* For any sub-region in a cluster $C_i$, which is composed of $n$ ($n \geq 1$) grid cells, $C_i.SGS$ can accurately express its density.

*Proof:* Since the skeletal grid cells in $C_i.SGS$ don't overlap, the population recorded by each skeletal grid cell accurately reflects the number of objects in it. Therefore, for any sub-region covered by the $n$ skeletal grid cells belonging to $C_i$, we can accurately calculate its density by dividing its total population by its total volume. ∎

**Lemma** 4.5. *Fidelity to Connectivity:* If there are two sub-regions in $C_i$ connected through a connected core object path composed of $n$ core objects, there must exist a core grid path connecting these two sub-regions with **at most** $n$ core cells on this path.

*Proof:* Since any skeletal grid cell containing a *core object* is a *core cell*, if there exists a *core object* path between two sub-regions, there must exist a *core cell* path between them. In the worst case, each *core grid* on this *core grid* path contains only one *core object*. Thus the length of the *core grid* path is at most equal to the length of the *core object* path. ∎

In conclusion, SGS effectively captures all key features of density-based clusters using a compact description.

## 5. PATTERN EXTRACTOR

Next, we introduce the pattern extractor that executes the Continuous Clustering Query (Section 3.2), outputting clusters in both *full* and *summarized (SGS) representations*. To provide such functionalities, a straightforward approach would be a two-stage process, namely cluster extraction followed by summarization. However, this strategy causes a significant performance overhead compared to cluster extraction only. An in-depth analysis of such a two-phase strategy can be found in our technical report [18].

### 5.1 Proposed Solution: Integrated Process

To solve this problem, we instead propose an integrated strategy that incorporates cluster extraction and summarization into a single process. The key observation that motivates this integrated computation method is given below.

**Observation** 5.1. *The main tasks for both density-based cluster extraction and SGS computation are the same, namely to first identify the connections (neighborships) among the objects and analyze them to form the cluster structures (in either the full or a summarized representation).*

This observation reveals the key commonality among the cluster extraction and summarization processes. Based on it, we design an integrated extraction+summarization method to effectively share the *neighborship* identification and cluster formation processes.

### 5.2 Incremental Computation and Challenges

To avoid conducting the prohibitively expensive clustering process from scratch at each window, our proposed method incrementally maintains the cluster structures across the windows. To realize incremental computation, we need to find an appropriate meta-data that can be maintained for both the full and summarized cluster representations. Our proposed solution is that, besides the raw data falling into each window, which needs to be maintained for cluster extraction in any case, we incrementally maintain the *skeletal grid cells* in the data space. With updated *skeletal grid cells*, we can easily output both the summarized and full representations of detected clusters. First, based on connections among the *skeletal grid cells*, we can easily determine the summarized representation SGS (a group of connected *skeletal grid cells*) for each cluster. Second, given the SGS of a cluster $C_i$, $C_i.SGS$, we can figure out the cluster member objects of $C_i$ based on the objects falling into the respective *skeletal grid cells* belonging to $C_i.SGS$.

However, incrementally maintaining *skeletal grid cells* in an efficient manner is a challenging task. In particular, tracking the changes to the *skeletal grid cells* caused by expired objects can be extremely expensive in terms of system resource utilization, and thus constitutes the key performance bottleneck for *skeletal grid cell* maintenance.

When an object $p_{exp}$ expires, it needs the connections at the object level, to update the connections among the *skeletal grid cells*. For example, when $p_{exp}$ expires, we first need to know which objects are neighbors of $p_{exp}$, as their *neighborships* with $p_{exp}$ will end from now on. This may break the connections between the *skeletal grid cell* $Sc_i$ in which $p_{new}$ resides and those in which $p_{exp}$'s neighbors reside. However, considering the large amount of pair-wise *neighborships* that may exist among the objects, maintaining all of them has been shown to be extremely expensive in terms of system resource utilization, analytically and experimentally [16]. Therefore, the straightforward incremental maintenance method, which updates *skeletal grid cells* corresponding to each insertion and deletion, is not practical.

### 5.3 "lifespan" Analysis

To solve this computation bottleneck, we present a *skeletal grid cell* maintenance method using "lifespan" analysis. This method elegantly eliminates the need for handling the impact of expired objects on the *skeletal grid cells*. The solution is based on the observation that in the sliding window semantics the lifespan of any object as well as the *neighborships* among objects are deterministic. Therefore, at the insertion stage, when we handle the impact of new objects on the *skeletal grid cells*, we take the lifespans of the objects into consideration. In particular, we pre-determine the changes that will happen to the *skeletal grid cells* when these objects expire later. Then at the expiration stage, no further update is needed to handle the impact of expired objects. Thus we avoid the bottleneck discussed above.

Among the five attributes of a *skeletal grid cell*, except *location* and *side length* that are fixed over time, the other three, namely *population*, *status* and *connections* are changing over time as the objects come and go with each window slide. The *population* of each *skeletal grid cell* is easily trackable with a simple object counter. Thus, we focus on the lifespan analysis of the *status* and the *connections*.

**Basics for lifespan Analysis.** First, we start with analyzing the lifespan of individual objects.

**Observation** 5.2. *Given the slide size $Q.slide$ of a query $Q$ and the starting time of the current window $W_n.T_{start}$, the **lifespan** of an object $p_i$ in $W_n$ with time stamp $p_i.T$ is $p_i.lifespan = \lceil \frac{p_i.T - W_n.T_{start}}{Q.slide} \rceil$, indicating that $p_i$ will participate in windows $W_n$ to $W_{n+p_i.lifespan-1}$.*

The number of windows that an object $p_i$ can survive in is determined by after how many window slides that $p'_is$ time stamp will still be greater than the starting time of the window. Based on the lifespan of individual objects, we analyze the lifespan of *neighborship* between two objects.

**Observation** 5.3. *Given two objects $p_i$ and $p_j$, the neighborship between them, $Neighbor(p_i, p_j)$ will hold for $Neighbor(p_i, p_j).lifespan = Min(p_i.lifespan, p_j.lifespan)$ windows, namely, it will exist in all windows from $W_n$ to $W_{n+Neighbor(p_i,p_j).lifespan-1}$ until either $p_i$ or $p_j$ expires.*

Based on these observations, we can further analyze the lifespan of different stages of an object's "career".

**Observation** 5.4. *Given an object $p_i$ and all its neighbors objects $p_{nb1}$ to $p_{nbk}$, the number of windows in which $p_i$ will be a core object $p_i.core\_lifespan = Min(p_i.lifespan, win\_\theta^c\_nei)$, with $win\_\theta^c\_nei$ the number of windows in which at least $\theta^c$ objects within $p_{nb1}$ to $p_{nbk}$ will participate. The number of windows in which $p_i$ will be edge object $p_i.edge\_lifespan = Min[p_i.lifespan - p_i.core\_lifespan, Max_{1\leq j \leq k}(p_{nb_j}.core\_lifespan)]$*

Basically, an object will be a *core object* in all the windows that it has at least $\theta^c$ neighbors. It will be an *edge object* when it *core object career* ends (no longer has enough neighbors) but at least one of its neighbors is still a *core object*.

**lifespan at Grid Cell Level.** To tackle *skeletal grid cell* maintenance, now we extend the concept of lifespan from the object level to the grid cell level. In particular, we analyze how the lifespan of objects, their *neighborships* and their *career* affects the lifespan of *skeletal grid cells*' *status* and *connections*. For each *skeletal grid cell* $Sc_i$, we maintain one lifespan indicator for $Sc_i.status$ and one for each $Sc_i.connections[i]$. Each lifespan indicates that, based on the objects in the current window, in how many future windows the value of this attribute will persist. These indicators will be updated as new objects arrive.

**Lemma** 5.1. **Status lifespan.** *Given a skeletal grid cell $Sc_i$, all the objects $p_0$ to $p_n$ in $Sc_i$, the number of windows in which $Sc_i$ will be a core cell $SG_i.core\_lifespan = Max_{0\leq i \leq n}(p_i.core\_lifespan)$*

Lemma 5.1 can be deduced by definition of a *core cell* (Def. 4.2). Namely, $Sc_i$ is a *core cell* if it contains at least one *core object*.

**Lemma** 5.2. **Connection lifespan.** *Given two skeletal grid cells $Sc_i$ and $Sc_j$, and all objects in $Sc_i$, $p_0^{sc_i}$ to $p_n^{sc_i}$, and all objects in $Sc_j$, $p_0^{sc_j}$ to $p_m^{sc_j}$, the number of windows in which $Sc_i$ and $Sc_j$ will be connected is defined as $Connection(Sc_i, Sc_j).lifespan = Max[Min(p_a^{sg_i}.core\_life - span, p_b^{sg_j}.core\_lifespan, Neighbor(p_a^{sg_i}, p_b^{sg_j}).lifespan)]$, $\forall a \in [0, n]$, $b \in [0, m]$.*

This indicates that two *skeletal grid cells* remain connected if at least one pair of *core objects*, each from one *skeletal grid cell*, are neighbors to each other.

**Auxiliary Meta-Data.** To insure that we only run one range query search (rqs) for each new object and never rerun rqs for existing objects, we maintain an auxiliary meta information for each object in the window. In particular, we maintain a "non-core-career neighbor list" for each object $p_i$ to store all $p_i$'s neighbors in its "non core career". For example, $p_i$ currently may have 100 neighbors. Based on the lifespan analysis, it will be a *core object* for 3 windows and then due to most of its neighbors expiring, it will become a *edge object* for 2 windows before expiration. In this case, the "non-core-career neighbor list" of $p_i$ only contains its neighbors in the last 2 windows of its lifespan, say 5 objects.

The "non-core-career-neighbors" of each object are maintained in a dynamic hash table. The hash table of each object $p_i$ is initialized to have $n$ buckets, with $n$ the number of windows that $p_i$ can survive. The hash key of the table is the number of windows that a neighbor object can survive. For example, when a data point $p_i$ finds a "non-core-career-neighbor" $p_j$, $p_j$ will be added to the $k^{th}$ bucket of the hash table, with $k$ the number of windows $p_j$ can still survive (if $k$ is larger than the number of buckets remained on $p_i$, $p_j$ is put in the last bucket). At each window slide, we can simply remove the whole first bucket of each remaining object, as all the neighbors in this bucket must be expired after the window slide.

The number of neighbors in such "non-core-career neighbor list" is bounded by the constant $\theta^c$. Namely an object can never have more than $\theta^c$ neighbors in its non-core career, otherwise it would instead be a *core object* in those windows. This theoretical bound guarantees the "lightness" of this auxiliary meta-data. Also, it provides all necessary access to the objects' neighbors needed in our cluster extraction process. It thus guarantees that we only run the minimum number of range query searches (one for each new object) during the clustering.

## 5.4 C-SGS Algorithm

We call our proposed algorithm based on the maintenance of skeletal grid cells and lifespan analysis C-SGS.

**Initialization.** For a continuous clustering query, at the initialization stage, C-SGS builds a grid-based index whose grid cell size is equal to the size of the finest *skeletal grid size* for this query (see Section 4). We assign to each grid cell in this index the same attributes as the *skeletal grid cells*, while we set their status to be *noise*, density to be "0", and connections to be all "false" initially.

**Handling Insertions.** For each new object $p_{new}$ inserted into the window, C-SGS first loads it into its corresponding *skeletal grid cell* based on its position in the data space. Then, we run a range query search for $p_{new}$ to identify $p_{new}$'s neighbors. Based on the lifespan of $p_{new}$ and its neighbors (Lemma 5.2), we can determine the lifespan of the *neighborships* among them (Lemma 5.3), as well as the lifespan of different stages of $p'_{new}s$ "career" (Lemma 5.4). Using this information, we can now update the *status* and *connections* of the *skeletal grid cells* in which $p_{new}$ falls into and in which its neighbors reside.

For **status** of *skeletal grid cells*, the insertion of a new object may only cause two types of changes. Namely, it may "promote" the *skeletal grid cells* to become *core cells* or "prolong" their *core cell lifespans*.

**status promotion:** A new object $p_{new}$ may promote the *skeletal grid cell* $Sc_i$ that it resides in to become a *core cell*, if it becomes the first *core object* in $Sc_i$. In this case, we set the *status* of $Sc_i$ to *core cell* and set its core cell lifespan equal to the core object lifespan of $p_{new}$. An example of this case is shown Case 1 of status promotion in Figure 6.

$p_{new}$ may also cause a *status* change of a *skeletal grid cell* by upgrading its non-core-object neighbors, which reside in these affected *skeletal grid cells*, to *core objects*. In this case, for each upgraded neighbor $p_{upg}$ of $p_{new}$, we first determine the lifespan of $p_{upg}$'s career by analyzing itself and its neighbors. As every $p_{upg}$ was a non-core object, the "non-core-career neighbor list" will help us to quickly access all its neighbors without running range query search again. Thus, we update the *status* of the *skeletal grid cells* in which $p_{upg}$ resides to *core cell* and set its core grid lifespan equal to the core object lifespan of $p_{upg}$. Correspondingly, the "non-core-career neighbor list" of each $p_{upg}$ also needs to be updated to exclude those objects that will only be neighbor of $p_{upg}$ in its core object career. An example of this case is shown in Case 2 of status promotion in Figure 6.

**status prolong:** A new object $p_{new}$ may prolong the core cell lifespan of the *skeletal grid cell* $Sc_i$ in which it resides, if $p'_{new}s$ core object lifespan is longer than that of any existing object in $Sc_i$. In this case, we set $Sc'_is$ core cell lifespan equal to the core object lifespan of $p_{new}$. An example of this case is shown in Case 1 of status prolong in Figure 6.

$p_{new}$ may also prolong the core cell lifespans of the *skeletal grid cells* by extending $p_{new}$'s neighbors' core object lifespan. For each $p_{new}$'s neighbor whose core object lifespan is extended because of $p_{new}$'s arrival, $p_{cole}$, we first determine how long its core object lifespan is extended, by analyzing it would have at least $\theta^c$ neighbors in how many more windows after $p_{new}$ joining its neighborhood. Then, we update the core cell lifespan of the *skeletal grid cell* in which each $p_{cole}$ resides to the core object lifespan of the corresponding $p_{cole}$, if the later is longer. An example of this case is shown in Case 2 of status promotion in Figure 6.

For **connections** of *skeletal grid cells*, the insertion of a new object may only cause two types of changes. Namely, it may build new connections between *skeletal grid cells* or prolong the lifespan of existing connections. The maintenance process of the **connections** follows the same principles used in **status** maintenance logics (details omitted here for space reasons but can be found in [18]).

**Handling Expirations.** By using the lifespan analysis technique introduced above, the impact to the *skeletal grid cells* that could be caused by expiring objects has been pre-handled when objects arrive. Therefore, no maintenance effort is needed for handling cluster structure changes when individual objects expire. After the window slides, the only update needed for the attributes of *skeletal grid cells* is to check whether the new window is out of the lifespans. If the new window is out of its *core cell lifespan*, its *status* needs to be set back to *edge cell*. If the new window is out of the lifespan of any of its connections, the corresponding connection needs to be set back to "false".

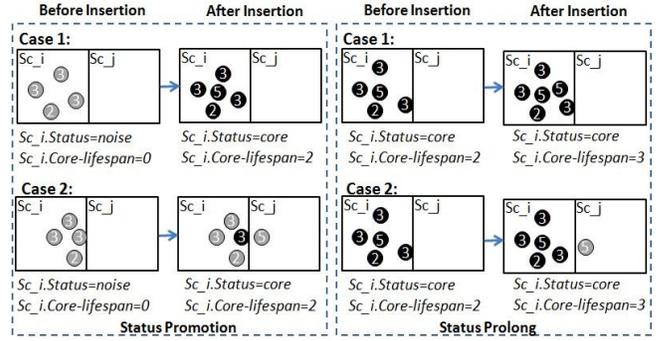

**Figure 6: Examples of updating cell status.** $\theta^c = 4$, grey circle=*edge point*, black circle=*core point*, number on each object= number of windows the object can survive.

**Output Stage.** At the output stage, the updated *skeletal grid cells* can be viewed as the vertices $V$ in a graph $G$, and the connections among them can be viewed as the edges $E$ among the vertices. Therefore, we simply conduct a depth first search on all the *core cells* to collect different groups of connected *core cells* and the *edge cells* attached to them. Each connected group of *skeletal grid cells* constitutes the SGS summarization of a cluster $C_i$, $C_i.SGS$. Given $C_i.SGS$, the *full representation* of $C_i$ can be easily figured out by collecting all objects covered by *core cells* in $C_i.SGS$ and those covered by the *edge cells* in $C_i.SGS$ and connected to at least one *core object* in $C_i.SGS's$ core cells.

## 6. PATTERN ARCHIVER

The pattern archiver handles two major tasks, namely pattern compression and selective pattern archival.

### 6.1 Multi-Resolution Cluster Summarization

Our proposed cluster summarization SGS supports multiple resolutions. In general, the SGS in different levels of resolution follow the same design as presented in Section 4. An SGS of any resolution is composed of a sequence of *skeletal grid cells*, and each *skeletal grid cell* has the same 5 attributes introduced before.

For any cluster $C_x$, the SGS of $C_x$ formed by the Pattern Extractor is based on the finest granularity, namely the smallest *skeletal grids cells*. Thus it is of the finest resolution. We call such $SGS$ the "*Basic SGS*" of $C_x$. The SGS in coarser resolutions are built based on hierarchically combining the *Basic SGS*. For a cluster $C_x$, we say that the *Basic SGS* of $C_x$ is at *Level 0* of the resolution hierarchy, noted as $C_x.SGS^{L_0}$. Any SGS in a coarser resolution is at a *Level n* denoted as $C_x.SGS^{L_n}$.

Each *skeletal grid cell* in $C_x.SGS^{L_n}$ ($n > 0$), $C_x.Sc_i^{L_n}$ is formed by combining the *skeletal grid cells* within a certain ($\theta$) sized hypercube space in $C_x.SGS^{L_{n-1}}$. For example, a 2-dimensional cluster $C_x$ has $SGS$ in two resolutions (Figure 5). They are at *Levels 0 and 1*. If the compression rate $\theta = 3$, each *skeletal grid cell* of *SGS* at *Level 0* is made by combining $3^2$ adjacent *skeletal grid cells* at *Level 1*. Both the number of resolutions allowed and the parameter $\theta$ are part of the configuration of our system.

Such compression process of building $C_x.SGS^{L_n}$ can be finished with a single scan of the *skeletal grid cells* in $C_x.SGS^{L_{n-1}}$. Given $C_x.SGS^{L_{n-1}}$ and to build $C_x.SGS^{L_n}$, we first generate a set of *skeletal grid cells* for $C_x.SGS_n^L$ to cover the whole data space occupied by corresponding cells in the $C_x.SGS^{L_{n-1}}$. Then we set the five attributes for $C_x.Sc_i^{L_n}$. The side length of any $C_x.SG_i^{L_n}$ is simply equal to the side length of a *skeletal grid cell* at *Level n-1* times $\theta$. Any $C_x.Sc_i^{L_n}$ is a *core cell* if at least one $C_x.Sc_i^{L_{n-1}}$ covered by it is a *core cell*. Otherwise, it is an *edge cell*. The population of any $C_x.Sc_i^{L_n}$ is equal to the sum of the population of the $C_x.Sc_i^{L_{n-1}}s$ covered by it. The connection vector of a $C_x.Sc_i^{L_n}$ is decided by the connections between the "boundary" $C_x.Sc^{L_{n-1}}s$ covered by it and those covered by its adjacent cells at *level n-1*.

**Budget- and Accuracy-Aware Resolution Selection.** Given the multiple resolution choices, the Pattern Archiver can decide in which resolution to archive the patterns based on both the system-resource budget and the accuracy required by the specific analytical tasks. In our SGS design, for a cluster summarization at a certain resolution, both its space consumption and conciseness are deterministic and easily calculatable. For space consumption, given the basic SGS of a cluster extracted, we can easily determine the number of *skeletal grid cells* needed in any other resolution for the same cluster, by calculating how many grid cells at that resolution are needed to cover the same data space. Since the *SGS* at different resolutions have the same design, the amount of information carried by each *skeletal grid cell* in any resolution is fixed. Thus, one can easily determine how much storage space is needed exactly for a given cluster in any resolution. For accuracy, as the size of the *skeletal grid cells* at all resolutions are known, the analysts would know exactly the granularity that their analytical task will be working on for a certain resolution.

## 6.2 Selective Pattern Archiving

The Pattern Archiver also selectively picks which clusters to archive. Currently, our system supports several simple but useful cluster selection mechanism, including using sampling techniques to select certain numbers of clusters to archive in a period of time and using feature selection to only archive clusters with certain features (e.g. only archive the clusters reaching a certain population or volume). More sophisticated pattern selection techniques, such as evolution driven techniques, will be studied in our future work.

## 7. PATTERN STORAGE AND MATCHING

## 7.1 Pattern Organization in Pattern Base

Our proposed cluster summarization method SGS empowers us to easily organize the extracted clusters based on their features. In particular, we build two indices for the archived clusters. One is based on the position of each cluster, and the second is based on all other features of each cluster captured in SGS.

We call the first index the *locational feature index*. As multi-dimensional objects, we express the position of each cluster using its minimum bounding rectangle (MBR). In our system, we employ one of the most widely used indices for MBRs, namely the R-tree index to organize them. The second index, called the *non-locational feature index*, organizes the clusters based on their non-locational features. We use a four-dimensional grid index to organize the clusters' SGS, with the four dimensions: the volume (number of *skeletal grid cells*, the status count (number of *core cells*), the average density and the average connectivity of each cluster.

## 7.2 Cluster Matching Process

The *Cluster Matching Queries* (see Figure 3) are executed by the Pattern Analyzer. To execute such queries, we first provide a distance metric (between 0-1) to measure the distance between two clusters. The metric is user-customizable based on application semantics.

$$Dist(C_a, C_b) = ps * Dist_{location} + \sum w_i * Dist_{nlf\_i}(C_a, C_b)$$

$$ps, Dist_{location} = 0 \| 1, \forall w_i, Dist_{nlf\_i} = [0, 1], \sum w_i = 1)$$

In this distance metric, $Dist_{location}$ indicates that whether two clusters overlap (1) or not (0). $ps$ indicates whether the matching is "position-senstive" (1) or not (0). $Dist_{nlf_i}$ represents the distance of two clusters on a specific non-locational feature and $w_i$ represents the analyst-specified weight on this feature.

To use this distance metric, the analyst needs to first specify whether the matching required by her application is position-sensitive, namely whether the matched clusters have to overlap in the data space. For the position-sensitive applications, $ps = 1$. If two clusters are not overlapped, $Dist_{location}(C_a, C_b) = 1$, the largest possible distance between two clusters, indicating that the two clusters are not similar and no further comparison on other features will be needed. For the non-position-sensitive applications, since $ps = 0$, the locational distance between two clusters is considered to be 0.

The second part of the distance metric measures the distance between two clusters on the four non-locational features, namely volume, status, population and connectivity. The distance on these features are used in both the match candidate search and detailed cell level cluster match.

**Candidate Search.** Given a to-be-matched cluster, a customized distance metric and a distance threshold specified by the analyst, our system first searches the potential match candidates in the Pattern Base. In the positional-sensitive case, the Pattern Analyzer first searches the *locational feature index* for the candidate clusters. If any overlapped clusters are found, it will calculate their non-locational distance with the to-be-matched clusters, and returns the similar clusters if the distances are smaller than the threshold. In the non-position-sensitive case, the Pattern Analyzer directly searches against the non-locational feature index for the candidates. Given the distance metric and the distance threshold, the Pattern Analyzer can determine the range of the search on each dimension (feature). For example, given the volume of the to-be-matched cluster equal to 20, the weight on size distance is 0.20, the overall distance threshold is 0.2, the volume of the candidate clusters have to be between 14 and 30. This is because any other number $x < 14 \| x > 30$ will make $abs(x-20)/min(x, 20) > (0.2/0.4)$, which will definitely not fulfill the search creteria. The same principle can be used on other features to determine the range of search. Given the search ranges on all dimensions, the Pattern Analyzer can quickly narrow down the candidate clusters to a small subset by searching the feature index.

**Grid Cell Level Cluster Match.** Given a to-be-matched cluster and a match candidate cluster for it, grid cell level cluster match compares the features of two clusters in their corresponding sub-regions (skeletal grid cells). In particular, grid cell level match uses the same customizable distance metric introduced earlier, while the distance between two clusters is now measured by aggregating the difference between all the corresponding *skeletal grid cell* pairs in these two clusters. More precisely, given a certain alignment between two clusters $C_a$ and $C_b$,[2] each *skeletal grid cell* $Sc_i$ in $C_a$ may have a corresponding *skeletal grid cell* in $Sc_j$, depending on whether its corresponding sub-region is also covered by $Sc_j$. If $Sc_i$ has a corresponding *skeletal grid cell* $Sc_j$ in $C_b$, their difference can be measured by comparing their status, density and connectivity features. Otherwise, $Sc_i$ is assigned the maximum difference with its corresponding sub-region, which is not a part of $C_b$ and thus can viewed as an empty grid. When calculating the distance between two clusters $C_a$ and $C_b$. we sum the difference between each $Sc_i$ in $C_a$ and its corresponding sub-region in $C_b$ to form the overall distance between the two clusters.

In the position-sensitive cases, no alignment is needed, or in other words, the alignment vector is always equal to [0,0,...,0]. This is because such applications require any *skeletal grid cell* $Sc_i$ in $C_a$ to be matched with the *skeletal grid cell* $Sc_j$ in $C_b$ that have the same absolute position in the data space. Therefore in such cases, we only need a single scan on the *skeletal grid cells* in two clusters to calculate the distances between them.

In the non-position-sensitive case, one or more alignments that minimize the distance between two clusters may exist. When given sufficient computation time, such as in an offline computation, one could apply an exhaustive search to find such an optimal alignment. In our system, for online computation, we use an A* style anytime search algorithm to search for the best alignment within a certain computation time budget. In particular, we start with an alignment that makes two clusters well overlapped. Then we continuously search along the direction of the most promising "nearby" alignment, which gives the smallest distance so far. When the given computation time budget is reached, we stop searching and return the smallest distance found so far as the distance between the two clusters.

## 8. EXPERIMENTAL EVALUATION

We conducted our experiments on a Dell desktop with an Intel Core2 2.2GHz processor and 3GB memory, running Windows 7 professional. We implemented the algorithms in VC++ 7.0.

**Real Datasets.** We used two real streaming datasets in our experiments. The first dataset, GMTI (Ground Moving Target Indicator) [6], records the real-time information on moving objects gathered by 24 different ground stations or aircraft in 6 hours from JointSTARS. It has around 100,000 records regarding the information on vehicles and helicopters (speed ranging from 0-200 mph) moving in a certain geographic region. The second real dataset we use is the Stock Trading Traces data (STT) from [11], which has one million transaction records throughout the trading hours of a day.

For the experiments that involve data sets larger than the sizes of these two datasets, we append multiple rounds of the original data varied by setting random differences on all attributes, until it reaches the desired size.

**Alternative Summarization Formats.** We compare our proposed Skeletal Grid Summarization (SGS) with three alternative cluster summarization formats. 1) The traditional "Centroid-Radius-Density" summarization (CRD). 2) Random Sampling Summarization (RSP). RSP for each cluster is generated by sampling the cluster members at a certain sampling rate $R$. To compare RSP with our proposed SGS summarization, for each specific cluster in the experiment, $R$ is always controlled to let its RSP have the same memory consumption with the SGS for the same cluster. 3) Skeletal Point Set (SkPS) summarization, our initial cluster summarization design proposed in Section 4.2.

### 8.1 Efficiency of Cluster Extraction + Summarization

In this experiment, we evaluate that how many system resources are needed to generate the alternative cluster summarizations respectively. Since our proposed solution, C-SGS, incorporates cluster extraction and summarization into a single process, we compare its performance with the following alternatives. 1) Extra-N: Extract clusters using state-of-the-art algorithm Extra-N [16] but do not generate any cluster summarization. 2) Extra-N + CRD: Extract clusters using Extra-N and then generate CRD for each extracted cluster. 3) Extra-N + RSP: Extract clusters using Extra-N and then generate RSP for each extracted cluster. 4) Extra-N + SkPS: Extract clusters using Extra-N algorithm and then generate (approximated) SkPS for each cluster using MG algorithm proposed in [9].

We first run each alternative method against the STT stream to extract clusters based on four dimensions, namely the transaction type (buy/sell), price, volume and time. To compare the performance of the alternatives when handling clusters with different characteristics, we use three different query parameter settings, namely case 1: ($\theta^r = 0.05$, $\theta^c = 10$), case 2: ($\theta^r = 0.1$, $\theta^c = 8$), case 3: ($\theta^r = 0.2$, $\theta^c = 5$). Also, for each case, we use three different window parameter settings, namely we fix the window size ($win$) for all three settings at 10K tuples, while varying the slide size $slide$ to equal to 0.1K, 1K and 5K tuples respectively.

For each case, we first verify the **correctness**[3] of our proposed C-SGS cluster extraction method by comparing the clusters extracted by it in full representation with those extracted by the state-of-the art technique Extra-N. In all the test cases, we found that the clusters extracted by C-SGS are identical with those extracted by Extra-N.

For **efficiency**, we measure two major performance metrics for stream processing: **1)** The average response time for each window (denoted as Response Time). For each window, we measure the average CPU time elapsed from the time that all new data have arrived to the time that all clusters have been output in both the full and summarized

---

[2] An alignment for two Skeletal Grid Summarizations (SGS) is a location shifting vector. For example, given two three dimensional clusters $C_a$ and $C_b$, an alignment equal to [1,2,1] indicates that any *skeletal grid cell* in $C_a$ with location vector equal to [x,y,z] corresponds to a *skeletal grid cell* in $C_b$ with location vector equal to [x+1,y+2,z+1], if any.

[3] All clustering algorithms following definition in [8] should produce the same clustering results given a same input object sequence.

130

representation. The average response time for each window shown in all cases are averaged among running for 10K windows. **2)** The memory footprint, namely the peak memory utilization of each alternative, among the 10K windows.

As shown in Figure 8.1, compared to Extra-N, which extracts clusters only but does not generate any cluster summarization (the baseline case), the other four alternatives, each generating a specific type of cluster summarization, exhibit some overheads in terms of CPU time utilization. However, such overhead caused by C-SGS, Extra-N + CRD, and Extra-N + RSP, is very modest, if not neglectable. The reason for such modest overhead caused by Extra-N + CRD and Extra-N + RSP is obvious. This is because CRD and RSP are very simple summarization formats that can easily be generated by at most two scans of the cluster members of each cluster. The overhead caused by our proposed solution C-SGS is comparable with those two simple summarization methods. This is because the major computation needed for generating the SGS cluster summarization, namely determining the status and connections among *skeletal grid cells*, is elegantly piggy-backed by the cluster extraction process itself. The CPU overhead of Extra-N + SkPS is significantly higher than that of the other alternatives. This is because generating SkPS is very expensive computationally [9]. For different window parameter settings, C-SGS has lower overhead for the settings with larger *win/slide* rates. This is because the performance of Extra-N is affected by the increasing number of "views" that needs to be maintained, which is equal to *win/slide* (see [16] for details), while the meta-data maintained by C-SGS and the corresponding maintenance effort is independent from this ratio.

Memory-wise, as shown in (Figure 8.1), our proposed method od C-SGS also exhibits very limited overhead in all test cases. This is because the process of generating SGS happens in place with the cluster extraction process.

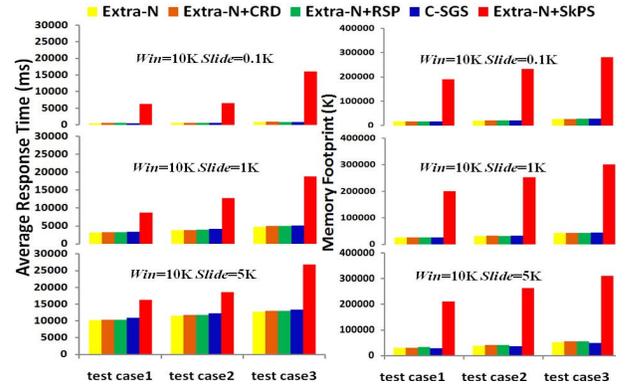

**Figure 7: CPU time and Memory comparisons for generating alternative summarizations.**

Similar performances are also observed in the same experiments but using GMTI data. We have also conducted an experiment showing the superiority of our proposed method when using time-based windows and under fluctuating input rate. The details of these experiments mentioned can be found in our technical report [18].

In conclusion, using our proposed C-SGS solution, we can efficiently generate the Skeletal Grid Summarization (SGS) for extracted clusters during online clustering process, with very limited system resource overhead.

## 8.2 Efficiency of Cluster Matching Queries

Next, we study the performance for running the cluster matching queries using our proposed summarization format SGS and other alternative summarization formats. We run three queries using the same pattern parameter settings as used in the previous experiment but with the same window parameter setting ($Win = 10K$, $Slide = 1K$) against the STT data using our proposed C-SGS method. We vary the size of the Pattern Base equal from 0.1K, 1K and 10K respectively. In each test case, we run each clustering query and archive all the clusters detected into the Pattern Base until the required number of archived clusters is reached. For each archived cluster, we also generate and keep the other three alternative cluster summarization formats for evaluating other matching methods. Once the required number of clusters is archived, we stop archiving and randomly pick 100 newly detected clusters as to-be-matched clusters. For each to-be-matched cluster, we run four matching queries for it against the archived clusters, each using one alternative cluster summarization method and the corresponding distance metric. In particular, we implement a subtraction function to measure the distance between the CRD of two clusters, which gives equal weight to the three cluster features captured in CRD, namely the centroid, range and density. We use the subset matching algorithm presented in [15] to calculate the distance between the RSP of two clusters. We use the graph edit distance algorithm presented in [13] to calculate the distance between the SkPS of two clusters. We give equal weight to all four features when measuring the distance between the SGSs of two clusters.

For each Pattern Base size, we measure the average response time for all cluster matching queries and memory space consumed by storing cluster summarizations.

As shown in Figure 8, when matching against 0.1K clusters, the average response time for each cluster matching query using SGS is less than 0.1 second. For the 1K and 10K cases, the average response time for our solution is only around 0.5 seconds and 3 seconds. Such high efficiency is comparable with cluster matching using CRD, which is very fast because of its extremely simple matching mechanism (simply three subtraction operations). This is due to the design of SGS, which effectively summarizes the key features of each cluster on both cluster and grid levels. In particular, by using our proposed two-phase matching strategy, the majority of the candidates in the pattern base are filtered out in the summarization matching phase. Thus, the more expensive grid level matching is only needed for a very small portion of the candidates. In our experiment, we found that only 6% of the candidate clusters necessitated the grid level match on average during the cluster matching process.

Memory-wise, SGS consumes only 0.12M, 1.38M and 12.24M memory space to store 0.1K, 1K and 10K clusters respectively (Figure 8). In particular, each 4-dimensional *skeletal grid cell* only consumes 23 bytes, position: 16 bytes (4 integers), status: 1 byte (1 boolean), density: 4 bytes (1 integer), connection: 2 bytes ($2^4 = 16$ booleans). In all test cases, the average number of *skeletal grid cells* in each cluster is 68. Therefore, only 1.5K memory is needed to store the SGS of each cluster on average. Compared with the memory space needed for storing the full representation of the clusters, which need 6.4M, 75.2M and 680.2M to store 0.1K, 1K and 10K clusters respectively, the average compression rate of SGS in our experiment is around 98%.

In conclusion, our proposed solution demonstrates high efficiency for cluster matching queries, which is significantly better than matching SkPS or RSP. Its performance is comparable with matching simple CRD cluster summarization. However, matching CRD is shown to have a much worse cluster matching quality compared with our proposed method of matching SGS (see next experiment below).

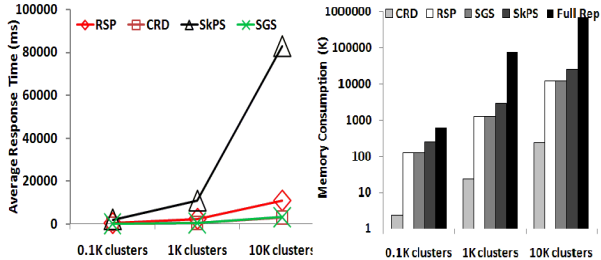

Figure 8: CPU time and Memory comparison for cluster matching queries using alternative cluster summarization methods

## 8.3 Quality of Cluster Matching

To measure the quality of cluster matching using alternative summarization formats, we invited 20 human analyts (all WPI graduate students) to visually analyze the similarity between the to-be-matched cluster and the matched clusters found for them using one alternative method. The analysis process is supported by ViStream [14], a freeware multivariate data visualization tool, which has been shown to be effective for helping human analysts to observe and understand multi-dimenstional clusters in streams. For each to-be-matched cluster, the analysts are asked to rate the top three similar clusters found by each summarization format into three categories, namely "very similar", "similar", and "not similar".

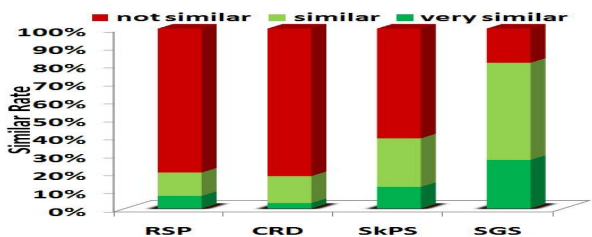

Figure 9: Similar rate given by users for matched clusters found by alternative summarizations

As shown in Figure 8.3, our proposed summarization method SGS demonstrates a high "similar rate", which is significantly better than all the other alternatives. This indicates that the human analysts agree with most of the similar clusters found using SGS, while disagreeing on a large percentage of those found using other alternatives. This shows the high effectiveness of SGS summarization in terms of cluster matching. Due to page limit, the detailed experimental setup and result analysis of this experiment are omitted here but can be found in our technical report [18].

We also conducted a series of experiments to confirm both the efficiency and effectiveness of cluster matching queries when using SGS with different resolutions. The details of those experiments can be found in our technical report [18].

## 9. CONCLUSION

In this work, we present a framework to support summarization and matching of density-based clusters in streaming environments. First, our work solves several open problems for density-based cluster analysis, namely, designing a descriptive yet compact summarization method for such clusters. Second, we present an efficient computation strategy to quickly summarize the detected clusters into SGS during the online clustering. Lastly, we design a cluster archiving and matching mechanism, which allows the analysts to submit cluster matching queries to find similar clusters detected earlier in the stream history. Our experimental study demonstrates the clear superiority of our proposed methods on both the efficiency and effectiveness.